\documentclass[prl,twocolumn,superscriptaddress,amsmath,amssymb,showpacs]{revtex4}

\newcommand{\be}{\begin{equation}}
\newcommand{\ee}{\end{equation}}
\newcommand{\ben}{\begin{eqnarray}}
\newcommand{\een}{\end{eqnarray}}
\newcommand{\bra}[1]{\langle #1|}
\newcommand{\ket}[1]{|#1\rangle}

\newcommand{\meanp}{\langle p\rangle}
\newcommand{\bsen}{\begin{subeqnarray}}
\newcommand{\esen}{\end{subeqnarray}} 

\usepackage{graphicx,graphics}
\usepackage{mathrsfs}
\usepackage{dcolumn}
\usepackage{bm}
\usepackage{subfigure}
\usepackage{amssymb}
\usepackage{epsfig}
\usepackage{amsfonts}
\usepackage{subeqnarray}
\usepackage{appendix}
\usepackage{srcltx}
\usepackage{color}
\usepackage[colorlinks=true, pdfstartview=FitV, linkcolor=red, citecolor=blue, urlcolor=blue]{hyperref}
\begin{document}

\title{Resonance Effects in the Nonadiabatic Nonlinear Quantum Dimer}
\author{Mukesh Tiwari} \email{mukesh\_tiwari@daiict.ac.in}
\affiliation{Dhirubhai Ambani Institute of Information \& Communication Technology (DA-IICT), Gandhinagar, 382007, India}
\affiliation{Consortium of the Americas for Interdisciplinary Sciences, University of New Mexico, Albuquerque, New Mexico, 87131-1156,USA}

\author{D.~V.~Seletskiy} \email{denisel@unm.edu}
\affiliation{Department of Physics \& Astronomy, University of New Mexico, Albuquerque, New Mexico, 87131-1156,USA}

\author{V.~M.~Kenkre} \email{kenkre@unm.edu}
\affiliation{Consortium of the Americas for Interdisciplinary Sciences, University of New Mexico, Albuquerque, New Mexico, 87131-1156,USA}
\affiliation{Department of Physics \& Astronomy, University of New Mexico, Albuquerque, New Mexico, 87131-1156,USA}

\begin{abstract}

	The  quantum nonlinear dimer consisting of an electron shuttling between the two sites and in \emph{weak} interaction with vibrations, is studied numerically under the application of a DC electric field. A field-induced resonance phenomenon between the vibrations and the electronic oscillations is found to influence the electronic transport greatly. For initially delocalization of the electron, the resonance has the effect of a dramatic increase in the transport. Nonlinear frequency mixing is identified as the main mechanism that influences transport. A characterization of the frequency spectrum is also presented.
\end{abstract}

\maketitle

\section{Introduction}

The response of an electron-phonon system to an externally applied electric field forms an important area of research in condensed matter physics \cite{{Kuehn2010},{Goychuk2005},{Janssen1995},{Meinert2001},{Banyai1993}}. For example, it is known that coupling between the electron and phonons leads to delocalization of the initially localized Wannier-Stark states under the application of a constant electric field (see for e.g. \cite{{Emin1987},{Hart1988}}), and a nontrivial response to a sinusoidally varying field applied to an electron in a lattice results in dynamic localization of the electron \cite{Dunlap1986,Raizan1998,KenRaghav2000,Kenkre2000}. By tuning the Wannier-Stark splitting to overlap with the lattice vibration frequency, new optical phonon mediated paths have been observed experimentally in superlattices \cite{Kast2002}. Advent of novel terahertz spectroscopy techniques has made it possible to observe nonlinear response of systems to applied electric fields in real time along with the small energy scales. Experimental observation of the evolution of long-range correlations leading to the birth of the quasiparticles \cite{Huber2001}, resolution of the dynamics of the nonlinear transport of polarons in the presence of strong quasi-stationary fields \cite{Gaal2007}, ultrafast resolution of the elementary excitations in superconductors \cite{Pashkin2010}, observation of the strong mode-coupling dynamics in the cavity quantum electrodynamics \cite{Gunter2009} for example, are some of the recent hallmark demonstrations of the modern experimental sophistication to resolve real-time response of the modes of the condensed matter systems to the external perturbations (e.g.~electric fields).
Motivated by these recent experimental progress, this paper numerically investigates electronic transport in a simple dimer system, consisting of an electron oscillating between two sites in the presence of the applied DC electric field. The electron, described quantum-mechanically, is allowed to  couple \emph{weakly} with the vibrational degree of freedom, represented by a classical simple harmonic oscillator. Henceforth in this paper we will refer to the oscillations of this simple oscillator as a phonon field or lattice vibrations.

Most generally, the system is described by a Hamiltonian that separates the electron phonon system of interest into three parts \cite{{Schanz1997},{Scott1982}},
\be\label{eq:totalham} 
H_{tot}= H_{el} + H_{ph} + H_{int}
\ee

\noindent
where the first term describes a tight-binding electron in a 1-dimensional lattice:
\be\label{eq:hamelectronic}
H_{el} = V\sum_{m}\ket{m}\bra{m+1} + \ket{m}\bra{m-1} + E \sum_{m}m \ket{m}\bra{m}\cdot
\ee

\noindent
with the lattice site integer $m$ running from $-\infty$ to $+\infty$, $V$ is the nearest neighbor transfer rate, and $E = e a\mathcal{E}$ is proportional to the magnitude of the degeneracy lifting electric field $\mathcal{E}$ with proportionality constant as a product of electronic charge $e$ and a lattice constant $a$. The phonon term in Eq.~(\ref{eq:totalham}) is represented by a collection of identical Einstein oscillators of mass $M$ and frequency $\omega_{0}$,
\be\label{eq:hamphononic}
H_{ph} = \frac{1}{2M}\sum_{m} p_{m}^{2} +\frac{M\omega_{0}^{2}}{2} \sum_{m} x_{m}^{2},
\ee
\noindent
where, $x_{m}$ and $p_{m}$ are respectively, the position and momentum of the $m$th oscillator. The interaction (Eq.~(\ref{eq:totalham})) between the electron and the lattice is taken to be linear in the displacement of the oscillator coordinate and to modulate only the site energy of the electron and not the intersite transfer matrix elements:
\be
H_{int} = \alpha\sum_{m}\ket{m}\bra{m}x_{m}.
\ee
\noindent
The parameter $\alpha$ describes the strength of the interaction. 

In this paper we focus on the problem in which the lattice is restricted to only 2 sites. The 2-site system, also known as the spin-boson system or the 2-site Holstein polaron \cite{{Lu2003},{Kenkre2004}}, has been explored extensively due to its formal simplicity and capability to reproduce certain features observed in extended systems rather accurately \cite{{Kenkre1986},{Kenkre1989}}. Despite the simplicity of the system, the apparent complexity in its behavior is evident from the fact that the complete quantum mechanical problem can not be solved exactly analytically. Perturbation methods and approximation schemes are thus commonly used to obtain insight into the system. In the present work we concentrate on the semiclassical approximation scheme  \cite{{Andersen1993},{Cruzeiro-Hansson1997},{Eilbeck1985},{Kenkre1994},{Kenkre1992}}, the range of validity for which was investigated elsewhere \cite{{Salkola1995},{Bonci1993}}. 

In the semiclassical approximation, where only the electron is treated quantum-mechanically, the equations of motion are derived by assuming for the electronic states $\ket{\psi_{e}}(t) = \sum_{m}c_{m}(t)\ket{m}$, where $c_{m}(t)$ is the probability amplitude to find the electron in the localized state $\ket{m}$. The time evolution of the probability amplitudes $c_{m}(t)$ are obtained from the Schr\"{o}dinger equation $\left(i\hbar~ d\ket{\psi_{e}}/dt = (H_{el} + H_{int})\ket{\psi_{e}}\right)$. For the equation of motion for the lattice one uses the classical Hamiltonian $\langle H\rangle = \bra{\psi_{e}}H_{tot}\ket{\psi_{e}}$, and the Hamilton's equations $\dot{x}_{n}= \partial \langle H\rangle/\partial p_{n}$ and $\dot{p}_{n} = -\partial \langle H \rangle/\partial x_{n}$. In such case, one obtains,
\ben\label{eq:newton_chain}\nonumber
i\hbar \dot{c}_{m}&=& V\left(c_{m+1} + c_{m-1}\right) + \alpha x_{m}c_{m}\\
\ddot{x}_{m} &=& -\omega_{0}^{2}x_{m} -\frac{\alpha}{M}\left|c_{m}\right|^{2}\cdot
\een

\noindent
The equations of motion governing the motion of the electron and lattice for a 2-site system are immediately obtained from Eqs.~(\ref{eq:newton_chain}) and are given by,
\bsen\label{eq:dimer}\slabel{eq:dimer1}
&i\hbar\dot{c}_{1}= V c_{2} + \alpha x_{1}c_{1} - E c_{1}&\\\slabel{eq:dimer2}
&i\hbar\dot{c}_{2}=  V c_{1} + \alpha x_{2}c_{2} + E c_{2}&\\\slabel{eq:dimer3}
&\ddot{x}_{1} + \omega_{0}^{2} x_{1} = -\frac{\alpha}{M}|c_{1}|^{2}&\\\slabel{eq:dimer4}
&\ddot{x}_{2} + \omega_{0}^{2} x_{2} = -\frac{\alpha}{M}|c_{2}|^{2}&
\esen


\noindent

Pauli spin matrices $\hat{\sigma}_{x},~\hat{\sigma}_{y},~\hat{\sigma}_{z}$ here $p,~q,~r$ respectively, can be obtained through 
$ p = c_{11} - c_{12},~q =  i(c_{12} - c_{21}),~r= c_{12} + c_{21}$ with $c_{ij} = c_{i}c_{j}^{*}$. Furthermore by defining a dimensionless internal coordinate $y =\frac{M\omega_{0}^{2}}{\alpha}\left(x_{2}-x_{1}\right)$ Eqs.~(\ref{eq:dimer}) can be 
recast in terms of real quantities only to give \cite{{Kus1994},{Kenkre1989}, {Kenkre2004}},
\bsen
\label{eq:solve}
\slabel{eq:solve1}
&\dot{p} = q&\\\slabel{eq:solve2}
&\dot{q} = -p - \chi y r - \Delta r&\\\slabel{eq:solve3}
&\dot{r}= \chi y q + \Delta q &\\\slabel{eq:solve4}
&\ddot{y} + \omega^{2}y = \omega^{2} p\cdot&
\esen

\noindent
where $\dot{p}$ implies a derivative with respect to a dimensionless time $\tau = \left(2V/\hbar\right)t$.
The various dimensionless parameters appearing in Eqs.~(\ref{eq:solve}) are $\chi = \alpha^{2}/\left(2M\omega_{0}^{2}V\right),~\Delta = E/V$ and $\omega = \hbar\omega_{0}/(2V)$. The main observable $-1\leq p \leq 1$ describes the difference in the probability of occupancy of the two sites of the dimer, while, $y$ is the scaled internal coordinate of the lattice displacement. The limit of weak coupling of electron to the lattice corresponds to $\chi\ll 1$ and we will restrict our analysis only to this regime.

Various variants of Eqs.~(\ref{eq:solve}) have appeared extensively in the literature. For example, under the adiabatic approximation $(y = p)$, Eqs.~(\ref{eq:solve}) can be written in a closed form in $p$ as,  
\be\label{eq:adiabatic}
\ddot{p} + \left(1 + \Delta^{2} + \chi c_{0}\right) p = -\Delta c_{0} - \frac{3\Delta\chi}{2}p^{2} - \frac{\chi^{2}}{2}p^{3}
\ee

\noindent
with $c_{0} = r_{0} - \frac{\chi p_{0}^{2}}{2} -\Delta p_{0}$ as a constant which depends only on the initial conditions and the system parameters. Eq.~(\ref{eq:adiabatic}) is similar to the equation for a degenerate trimer the exact solution to which in the form of Weierstrass Elliptic functions has already been obtained \cite{Andersen1993a}. Setting $\Delta = 0$ in Eq.~(\ref{eq:adiabatic}) reduces it to the well known Discrete Nonlinear Schr\"{o}dinger equation (DNLSE), for which the exact solution is in the form of Elliptic functions \cite{Kenkre1986}. Amongst other important studies the response to different initial conditions \cite{{Tsironis1988},{{Raghavan1997}}}, calculation of line shape in neutron -scattering \cite{Kenkre1987} etc. have been studied. Attempts have also been made to address the issue of validity of the adiabatic approximation for a degenerate dimer. Eliminating the approximation renders the problem extremely complicated, making it difficult to obtain analytical solutions. In certain instances, some analytical progress has been made either by introducing dissipation \cite{Kenkre1989}, or obtaining exact solutions for the non dissipative case, valid, for a limited set of initial conditions and parameter values \cite{Kus1994}. Our goal in this paper is two-fold: firstly, we do allow the lattice to vibrate on its own time scale and secondly, we focus on the response of the electron in the non adiabatic dimer to the application of electric field concentrating specifically on its transport.

\section{Results}

To quantify electronic transport in the dimer model of interest, it is important to identify time-independent (steady-state) parameters describing the dynamics of the system.  The importance of time averaged probability difference $\meanp = \lim_{T\rightarrow\infty} 1/T\int_{0}^{T}~d\tau~p(\tau)$ has been realized previously \cite{Andersen1993a}.  While $\meanp$ is a useful quantity in describing characteristics of the dynamics, it does not generally yield information about the transport without additional knowledge of the magnitude of the electronic oscillations. In this spirit, we introduce a quantity $\gamma_{d}$ given by
\be
\gamma_{d} =\left\{
\begin{array}{ll}
\left(1 - p_{max}/\meanp\right) , & \meanp < 0, \\ 
\infty , & \meanp = 0 , \\ 
1 - p_{min}/\meanp , & \meanp > 0,
\end{array}
\right.  \label{eq:deg_trapping}
\ee

\noindent
where $p_{max(min)}$ is the maximum (minimum) value of the probability difference $p$. Note that  $0\leq \gamma_{d}\leq 1$ implies localized dynamics while $\gamma_{d} \rightarrow \infty$ together with $p_{max(min)}\neq 0$ implies delocalized dynamics. We therefore refer to $\gamma_{d}$ as the degree of transport, which is also characteristic of the electrons mobility in this context.

To demonstrate the usefulness of the $\gamma_{d}$ measure, let us first turn off the lattice coupling by setting $\chi=0$ in Eq.~(\ref{eq:adiabatic}). Application of the electric field beyond a certain value would lead to the localization of the electron in the Wannier Stark states, and hence, to hindered electronic transport. This can be understood simply by considering a constant energy picture obtained from Eq.~(\ref{eq:adiabatic})
\be\label{eq:stark}
\frac{\dot{p}^{2}}{2} + U(p) = E_{constant}
\ee 

\noindent
where potential energy $U(p)= \frac{\omega_{e}^2}{2}p^{2} + \Delta c_{0}p$, with $\omega_{e} = \sqrt{1 + \Delta^2}$ being the bare-electron frequency. For asymmetric initial condition $\left(p_{0} = 1\right)$, $\meanp$ and $p_{min}$ can be trivially calculated from Eq.~(\ref{eq:stark}) to be $\frac{\Delta^{2}}{1 + \Delta^{2}}$ and $\frac{\Delta^{2}-1}{\Delta^{2} + 1}$ respectively, which when substituted in Eq.~(\ref{eq:deg_trapping}) gives $\gamma_{d} = \frac{1}{\Delta^{2}}$. Thus, $\gamma_{d}$ in the Stark case lies in the range $[0,\infty)$ depending upon the field strength: for $\Delta\rightarrow0$, $\gamma_{d} \rightarrow \infty$ showing complete transfer of the particle on both lattice sites with the probability difference oscillating between the two possible extrema $\pm 1$. Application of field results in a nonlinear reduction of the degree of transport with increasing field. $\Delta=1$ results in the localization of the electron on one of the lattice sites. Further increase in field value reduces the amplitude of oscillation and is effectively captured by $\gamma_{d}$. In the other extreme if we start with completely delocalized initial condition $p_0 = 0,~r_0=1$ which gives $\meanp = -\frac{\Delta}{1 + \Delta^{2}}$ and $p_{max} = 0$, for which $\gamma_{d} = 1$ independent of the value of $\Delta$. While $\gamma_{d}$ correctly predicts inhibited transport $(\text{particle now oscillates only on one site} )$, it does not capture the variation in the amplitude of oscillation as it is insensitive to the magnitude of the applied field. Similarly, in the degenerate adiabatic non linear dimer initially localized conditions results in the trapping of the electron on one lattice site as coupling parameter $\chi$ equals 2  \cite{Kenkre1986}. Further increase in field values lead to reduction in amplitude of the oscillation and is also captured by $\gamma_{d}$ which equals $\frac{\chi -\sqrt{\chi^2-4}}{\chi +\sqrt{\chi^2-4}}$ for $\chi\geq 2$. The other extreme with $r_0=1$ is a stationary solution of the problem leading to no oscillation in $p$. Breaking the degeneracy for this initial condition results in the confinement of the particle on single lattice site. It should be noted that if the lattice coordinate is not assigned any degree of freedom then completely delocalized initial condition leads to either confinement on one lattice site $\left(\Delta >0,~\chi=0\right)$, or stationary solution $\left(\Delta=0,~\chi>0\right)$ involving no oscillations in $p$. Even for $\Delta >0,~\chi>0$, this initial condition simply leads to reduction in amplitude of oscillation apart from confinement. This should also be evident from Eq.~(\ref{eq:adiabatic}) which in the potential picture represents the motion of a particle in a fictitious potential given by $U(p)=\Delta c_{0} p + \frac{\left(1 + \Delta^{2} + \chi c_{0}\right)}{2}p^{2} + \frac{\Delta\chi}{2}p^{3} + \frac{\chi^{2}}{8}p^{4}$. Increase in $\chi$ or $\Delta$ has the effect of increasing the potential barrier for the electron.

We now proceed to analyze the effect that applied electric field has on a quantum dimer with the lattice allowed to vibrate at its own characteristic frequency. For this, we consider the case of delocalized electronic initial conditions $(r_{0}=1)$ which corresponds to a trapped state in the adiabatic case and a stationary state for the degenerate non-adiabatic case. We keep the lattice to be initially unexcited $(\dot{y} = 0,~y=0)$ and stay in a regime in which the coupling between the lattice and the electron is weak $\left(\chi<<1\right)$.

\begin{figure}[!h]
\centering
\includegraphics[viewport = 80 220 580 620, width = 65 mm]{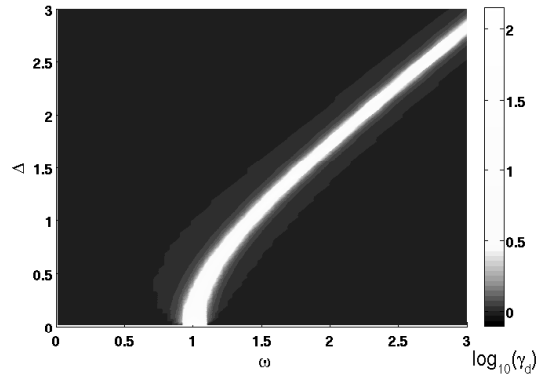}
\caption{A surface plot of the mobility of the quantum particle $\gamma_{d}(\Delta,\omega)$ for $\chi = 0.005$ and initially delocalized conditions $r_0 = 1, p_{0} = q_{0} = y_{0} = \dot{y}_{0} = 0$. A sharp increase  in mobility (white region) is observed along $\sqrt{1 + \Delta^{2}} \approx \omega$.}\label{fig:gdchip005}
\end{figure}

We investigate the behavior of the system numerically by analyzing the dependence of degree of transport $\gamma_d$ with varying electric field $\Delta$ and lattice frequency $\omega$. In Fig.~\ref{fig:gdchip005} we show a contour plot of $\gamma_{d}\left(\Delta,\omega\right)$ via simulations of Eqs.~(\ref{eq:solve}) for the above mentioned initial conditions. First, given field value $\Delta$ (y-axis) determines the bare-electron frequency $\omega_{e}^2 = 1  +\Delta^2$. By scanning the lattice frequency $\omega$ (x-axis), the behavior in adiabatic ($\omega_{e} \ll \omega$) and uncoupled-lattice ($\omega_{e} \gg \omega$) regimes can be observed.  Indeed, both regions show inhibited transport as $\gamma_{d} = 1$(Fig.~\ref{fig:gdchip005}, black color), consistent with the corresponding analytic results discussed above. Due to the presence of the interaction term $\chi$, the behavior of the system is however non-trivial when the two frequencies are in close proximity (Fig.~\ref{fig:gdchip005}, white color) resulting in strong enhancement of the electronic transport.  In this region both oscillators are resonantly coupled, exhibiting coupled mode behavior. In other words, the bare-electron frequency gets re-normalized upon being dressed by the interaction with the lattice.

\begin{figure}[!h]
\includegraphics[viewport = 24 182 690 580, width = 90 mm]{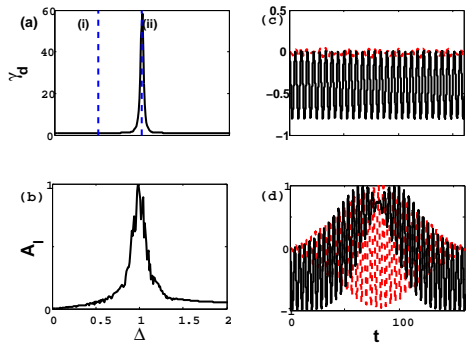}
\caption{Resonant enhancement in $\gamma_{d}$ (a) and normalized lattice amplitude $A_{l}$ (b) with varying electric field $\Delta$ for $\omega = \sqrt{2}$, $\chi$ = 0.005 and same initial conditions as in Fig.~\ref{fig:gdchip005}. Time evolution of $p$ (solid line) and oscillator coordinate $y$ (dashed line, normalized to resonant value in (b)) plotted for: (c) $\Delta  = 0.5$ away from resonance (vertical line labeled (i) in (a)); (d) $\Delta = 1$ (label (ii) in (a)) for one oscillation period of the difference frequency $\omega_{\text{DFG}}$.}
\label{fig:nlresponse}
\end{figure}

This point is further made clear in Fig.~\ref{fig:nlresponse} where the time evolution of both $p$ (solid line) and $y$ (dashed line) is plotted for cases (i) ``away'' (Fig.~\ref{fig:nlresponse}~(c)) and (ii) ``at'' (Fig.~\ref{fig:nlresponse}~(d)) the condition of resonance, as represented by the dashed vertical lines in Fig.~\ref{fig:nlresponse}~(a).  While on resonance, energy of electron motion gets transferred into the lattice mode, as shown by substantial increase in lattice oscillator amplitude $A_{l}$ (Fig.~\ref{fig:nlresponse}~(b)).  Lattice motion in turn leads to qualitative change of the electronic motion, resulting in un-trapped oscillations of the electron between two extrema of probability difference $p$, and hence enhanced transport. 

\begin{figure}[!h]
\centering
\includegraphics[viewport = 120 170 510 650, width = 60 mm]{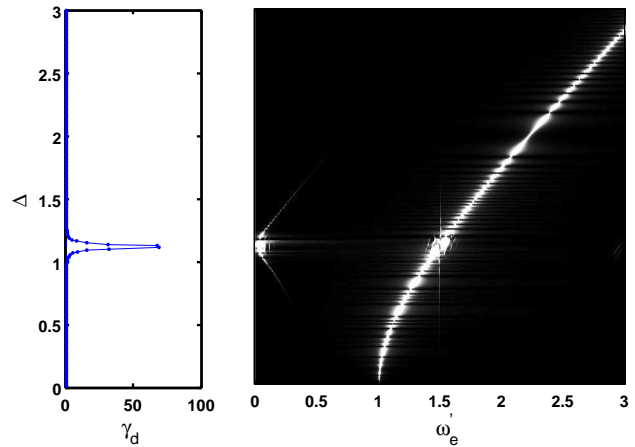}
\caption{Spectral response of the probability difference as a function of the electric field (right). The value of the coupling $\chi$ is $0.005$ while the vibrational frequency of the lattice is fixed at $\sqrt{2}$. On the left is shown the degree of transport $(\gamma_{d})$ with respect to the field. The dashed line at $\gamma_{d} = 1$ shows the absence of transport in the DC Stark case. Other parameters are the same as in Fig.~\ref{fig:gdchip005}.}\label{fig:fr_chip005}
\end{figure}

To understand the nature of the coupling we consider the behavior of the system in the frequency domain. To this end, we obtain frequency dependent probability difference $\mathcal{P}(\omega_{e})$ by taking the real part of the Fourier transform $\mathcal{P}(\omega_{e}) = \mathcal{R}\mathcal{F}[p(\tau)]$ of the time dependent solution $p(\tau)$. In Fig.~\ref{fig:fr_chip005} (right) we show the contour plot of $\mathcal{P}(\omega_{e},\Delta)$ for a fixed lattice frequency. As has been discussed above and is evident from $\gamma_{d}(\Delta)$ plot $(\text{Fig.~\ref{fig:fr_chip005}(left)})$ transport is enhanced \emph{only} in the region of coupling i.e. when the two frequencies $\omega_{e}$ and $\omega$ are in close proximity. What is evident from the contour plot, however, is also the presence of new frequency components which result from the nonlinear mixing of two fundamental frequencies $(\omega_{e}~\text{and}~\omega)$. New low frequency components occur when the two frequencies are at resonance, or in other words, the coupled modes are near the anti crossing point. It will be shown in the next section that the low frequency contribution is due to second order nonlinearity, resulting in the difference frequency generation (DFG): $\omega_{\text{DFG}} = \omega_{e} -\omega$, which approaches but never touches zero due to the anti crossing nature of the coupled modes.  It is this difference frequency term that is responsible for the enhancement of the transport on resonance, as this is the frequency with which electron tunnels between the dimer sites, assisted by the phonon field.

\section*{Characterization of Frequencies}

In the previous section we have seen that as the resonance condition is met, it results in the generation of higher order frequency components, in particular, the difference frequency component $\omega_{\text{DFG}}$ (see Fig.~\ref{fig:fr_chip005}). The new frequency components will in general carry signatures of the nature of coupling between the lattice and the electron present as a result of the linear interaction in the Hamiltonian (Eq.~(\ref{eq:hamelectronic})) and the semiclassical approximation resulting in Eqs.~(\ref{eq:solve}). To characterize the different frequency components present we first notice that starting from Eqs.~(\ref{eq:solve}) two integrals of motion can be derived \cite{Kus1994},
\bsen\label{eq:conservedqts}\slabel{eq:probconserved}
p^{2} + q^{2} + r^{2} &=& 1,~and \\\slabel{eq:Hamiltonian}
\frac{\dot{y}^{2}}{2\omega^{2}} + \frac{y^{2}}{2} - py + \frac{r}{\chi} - \frac{\Delta}{\chi}p &=& \mathcal{I}\cdot
\esen

\noindent
where Eq.~(\ref{eq:probconserved}) simply means that the total probability is always conserved, whereas Eq.~(\ref{eq:Hamiltonian}) is the statement of conservation of energy in the system. The presence of these two integrals of motion in Eqs.~(\ref{eq:conservedqts}), reduces the number of independent degrees of freedom to three, which still does not permit exact solutions. Solving for $r$ in Eq.~(\ref{eq:Hamiltonian}) and substituting in Eqs.~(\ref{eq:solve}), we obtain the following second order equation in $p$
\ben\label{eq:psecondorder}\nonumber
\ddot{p} =&-p\left[1 + \Delta^{2}\left\{1 + \epsilon y\right\}^{2}\right] -\Delta\left(1 +  \epsilon y\right)\chi\mathcal{I}& \\ &+ \frac{\chi\Delta}{2}\left(1 + \epsilon y\right)\left(y^{2} + \frac{\dot{y}^{2}}{\omega^{2}}\right),&
\een

\noindent
with $\epsilon$ as the ratio of interaction strength $\chi$ and the field strength $\Delta$. For situations in which the applied field is much stronger than the interaction strength we set $\epsilon = 0$ in Eq.~(\ref{eq:psecondorder}) to obtain,
\ben\label{eq:coupledosc}
\ddot{p} + \Big(1 + \Delta^{2}\Big)p&=& -\Delta\chi\mathcal{I} + \frac{\chi\Delta}{2}\left(y^{2} + \frac{\dot{y}^{2}}{\omega^{2}}\right)
\een

\noindent
which for our choice of initial conditions $\left(y_0 = \dot{y}_{0} = \dot{q}_{0}= 0\right)$ and by substituting $\mathcal{I} = r_{0}/\chi - \Delta p_0/\chi$ reduces to
\be\label{eq:reducedeqnp}
\ddot{p} + \Big(1 + \Delta^{2}\Big)p =  -\Delta\left(r_{0} - \Delta p_{0}\right) + \frac{\chi\Delta}{2}\left(y^{2} + \frac{\dot{y}^{2}}{\omega^{2}}\right)\cdot
\ee

\begin{figure}[!h]
\includegraphics[viewport = 54 182 559 610, width = 70 mm]{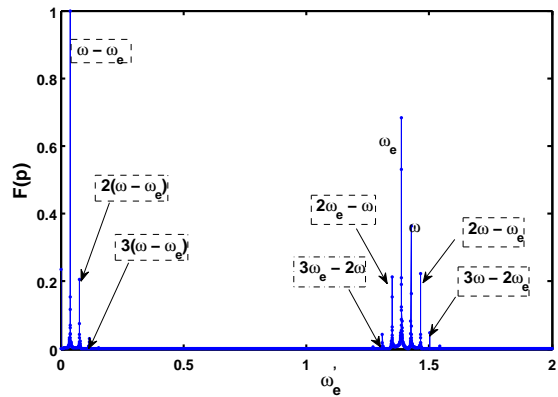}
\caption{Frequency spectrum of the non degenerate non-adiabatic dimer normalized to the maximum close to resonance. The different parameter values are $\chi =0.005$, $\Delta = 0.98$. The initial conditions are the same as in Fig.~\ref{fig:gdchip005}.}
\label{fig:fourierchidelta}
\end{figure}

\noindent
The set of equations given by Eq.~(\ref{eq:reducedeqnp}) and Eq.~(\ref{eq:solve4}) are able to reproduce the dynamics of the original equations (Eqs.~(\ref{eq:solve})) rather accurately for small values of $\chi$. Hence, in order to obtain an estimate of the different frequency components generated, instead of solving Eqs.~(\ref{eq:solve}), we take Eq.~(\ref{eq:reducedeqnp}) and Eq.~(\ref{eq:solve4}) as the point of departure. It is clear that the two fundamental frequencies in equations (\ref{eq:reducedeqnp}) and (\ref{eq:solve4}) are given by $\omega_{e}  = \sqrt{1 + \Delta^{2}}$ for the electron and $\omega$ for the lattice parameter $y$. Using $\alpha = \chi\Delta/2$ as the perturbation parameter we look for asymptotic solutions in the form of, 
\bsen\label{eq:pert}
p &=& \sum_{k}\alpha^{k}p_{k}\\
y &=& \sum_{k}\alpha^{k}y_{k}
\esen
where $k$ is the index of the power series expansion and takes values $0,1,2..\text{etc}$. Substituting Eqs.~(\ref{eq:pert}) in Eq.~(\ref{eq:reducedeqnp}) and Eq.~(\ref{eq:solve4}) and solving for each term of the perturbation series one can immediately see that the zero order term in $k$ is simply the DC Stark case of Eq.~(\ref{eq:adiabatic}) reproducing the stark frequency $\omega_{e} = \sqrt{1+ \Delta^{2}}$. $k=1$ term leads to the generation of new frequencies due to the possible two-particle interactions, i.e. self interaction resulting in second harmonic generation at 2$\omega_{e}$ and 2$\omega$ as well as interactions due to coupling of electron and lattice at the difference $\omega_{\text{DFG}} = \omega_{e} -\omega$ and sum $\omega_{\text{SFG}} = \omega_{e} +\omega$ frequencies. For $k=2$, in addition to the frequency components already present for $k = 0~\text{and}~1$ three-particle interactions give rise to frequencies at  $3\omega_{i,j}$, $|2\omega_{i} \pm \omega_{j}|$ where $i$ and $j$ represent electron and lattice interchangeably.  This behavior is verified by Fourier transforming numerical solutions of Eqs.~(\ref{eq:solve}) and is shown in Fig.~\ref{fig:fourierchidelta} for the case when electron is driven close to the lattice resonance.   In addition to the frequencies of two- and three-particle contributions analyzed above, higher order contributions of $4,~5$ etc. particle interactions is evident. As mentioned in the previous section, there is a strong difference frequency component $\omega_{\text{DFG}}$ that is responsible for the transport enhancement in the weak-coupling approximation ($\chi \ll 1$).  Finally, we want to point out that, as the resonance condition is approached, the approximation $\epsilon = 0$ fails. The qualitative effect of the field on the transport remains the same, but the fundamental and higher order electronic frequencies will depart from the values stated above.

\begin{figure}[!h]
\includegraphics[viewport = 54 182 559 610, width = 70 mm]{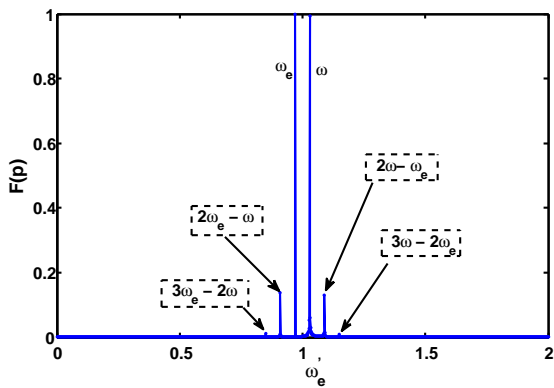}
\caption{Frequency spectrum of the degenerate non-adiabatic dimer at resonance from Eqs.~(\ref{eq:solve}) for localized initial condition $(p_{0} = 1)$. Other initial conditions are $q_{0} = r_{0} = y_{0} = \dot{y}_{0} = 0$. The coupling parameter $\chi = 0.005$. The two fundamental frequencies are denoted by $\omega_{e}$ and $\omega$.}
\label{fig:frresponsechi}
\end{figure}

So far we have shown that the application of the electric field results in transport enhancement through the generation of low-frequency $\omega_{\text{DFG}}$ component, where in via a two-particle process the electron is able to resonantly tunnel between the dimer sites, assisted by the interaction with a phonon.  To underscore the role of the applied field in the generation of the transport enhancement let us consider the case of a degenerate dimer near resonance, i.e. when $\omega_{e} \sim \omega=1$ (Fig.~\ref{fig:frresponsechi}) without the applied electric field. Since the delocalized initial condition corresponds to a stationary (non-oscillating) solution (as discussed in the previous section), we now consider an initially localized electron, i.e. $p_{0} = 1$, still in the weak-coupling regime ($\chi = 0.005$). In this case (see Fig.~\ref{fig:frresponsechi}) electron oscillations are supported at new frequencies resulting from the nonlinear mixing of the odd pairing of the two fundamental frequencies $\omega_{e}$ and $\omega$.  These new frequencies are namely $2\omega - \omega_{e}$, $2\omega_{e} -\omega$, $3\omega_{e} -2\omega$ along with the corresponding sum frequencies (not shown in Fig.~\ref{fig:frresponsechi}). The frequencies corresponding to even pairing (e.~g.~ $\omega_{\text{DFG}}$) are completely absent.  This is also evident from Eqs.~(\ref{eq:coupledosc}), which, for the case of $\Delta = 0$:
\be\label{eq:pchi}
\ddot{p} + \left(1 + \chi^{2}y^{2}\right)p = -\chi y r_{0} + 
\frac{\chi^{2}y}{2}\left(y^{2} + \frac{\dot{y}^{2}}{\omega^{2}}\right)\cdot
\ee

\noindent
As before, the exact solution to  Eq.~(\ref{eq:pchi}) and Eq.~(\ref{eq:solve4}) is not possible.  Following the perturbative analysis, the zeroth order term in $p$ represents the equation of a simple harmonic oscillator with no coupling to the lattice which has non zero solutions for not completely delocalized initial conditions $(r_{0}\neq 0)$. The first order correction represents the equation of two linearly coupled harmonic oscillators while the nonlinear mixing of frequencies can only be observed in the second and higher order terms.  The newly generated frequencies result from the odd mixing of the two fundamental frequencies due to the cubic variation with respect to the oscillator variables. This situation should be contrasted with the case of non-zero electric field, presented by Eq.~(\ref{eq:reducedeqnp}). There, the dependence on the lattice variables is quadratic, which means that while Eq.~(\ref{eq:pchi}) and Eq.~(\ref{eq:solve4}) are symmetric under the transformations  $p\rightarrow -p$ and $y\rightarrow -y$ Eq.~(\ref{eq:reducedeqnp}) and Eq.~(\ref{eq:solve4}) are not. The fact that the potential remains centrosymmetric without the applied electric field prohibits the generation of the even harmonics.  Application of the electric field results in the introduction of non-centrosymmetric component of the potential, hence allowing for even harmonics to be generated.  


The impact of different initial conditions on electronic transport can also be understood from Eqs.~(\ref{eq:reducedeqnp}) and(or) from Eqs.~(\ref{eq:Hamiltonian}). $p_{0}=1$ initial condition corresponds to solutions in which the electron and lattice oscillations are completely out of phase, whereas $r_{0}=1$ corresponds to the symmetric mode or in phase oscillation of the two oscillating systems. Completely delocalized initial condition as seen in the Stark case implies enhanced tunneling. These two effects coupled with the nonlinear interaction makes this initial condition more favorable for enhancement of transport.

\section{Conclusions}

In this paper we have presented a study of the response of the nonlinear quantum dimer to a constant externally applied electric field. Our studies differ from the previous ones (see for e.g. \cite{Kenkre1986, Kenkre1989}) in that we allow the lattice to evolve dynamically at its own time scale. Inclusion of the lattice degree of freedom results in a resonant enhancement in the transport of the electron. We performed numerical and perturbation series analysis in the limit of weak coupling of the electron with the lattice ($\chi \ll 1$) to provide a deeper insight into the ongoing processes.  We find that application of the DC electric field induces a second-order nonlinear mixing of the electron ($\omega_{e}$) and lattice ($\omega$) frequencies facilitating electronic transport enhancement near resonance. This mixing is manifested by a two-particle interaction resulting in a strong difference frequency component $\omega_{DFG} = \omega_{e} - \omega$, which corresponds to the oscillation of the slowly varying envelope of $p$ (see Fig.~\ref{fig:nlresponse}). By symmetry arguments, $\omega_{DFG}$ arises due to the applied field and would not have been present, otherwise. The observed enhancement of the transport is similar to the phonon-assisted electron hopping, well-known for extended systems \cite{Emin1987}.  Despite the classical treatment of the lattice, the subtraction of the lattice frequency $\omega$ in the DFG process is akin to single phonon annihilation. This means that the electron absorbs one quantum of the lattice energy to complete the transfer to the next lattice site.     

\begin{figure}[!h]
\centering
\includegraphics[viewport = 120 180 540 610, width = 65 mm]{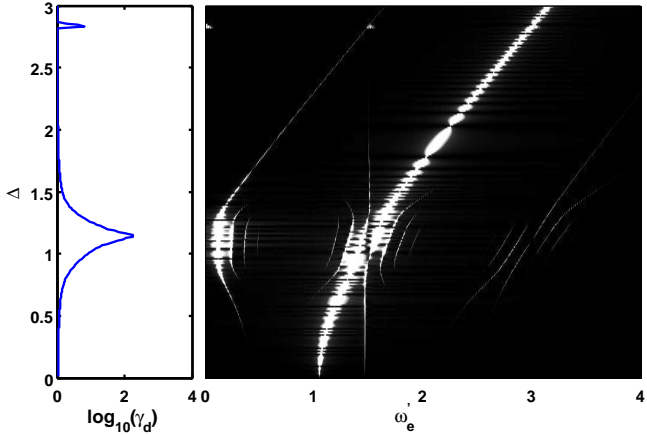}
\caption{Spectral response for the probability difference as a function of the electric field (right). The value of the coupling $\chi$ and lattice vibrational frequency $\omega$ was taken to be $0.05$ and $1.5$ respectively. On the left is shown the degree of transport $\gamma_{d}$ corresponding to that value of the field on a logarithmic scale. Other parameters are the same as in Fig.~\ref{fig:gdchip005}.}\label{fig:chip05_freq}
\end{figure}

We would like to further strengthen this last point by demonstrating the behavior of the system under the application of DC field with increased strength of coupling (Fig.~\ref{fig:chip05_freq}). We show the variation in degree of transport with changing field (left), on a logarithmic scale and the contour plot of the different frequency components generated. Apart from the already discussed resonant enhancement $\omega_{e} \approx \omega$, a sharp increase in $\gamma_{d}$ is also seen at a higher field value $(\Delta\approx 2.83)$, accompanied by a low-frequency contribution as depicted on the contour plot. This enhancement can be assigned to the field-induced three-particle process in which electron transfer to the next dimer site is assisted by the annihilation of the two quanta of the lattice vibration.  Due to the higher order nonlinear process, the linewidth of the enhancement $\gamma_{d}(\Delta)$ is narrower than in the single phonon interaction. For stronger $\chi$ values, the enhanced transport is seen at a multitude of increasing field values, corresponding to the generation of low frequency components. In such cases, system quickly leaves perturbative regime and becomes chaotic, making it difficult to obtain a clear insight into the phenomenon. 

Our study should be relevant to systems which are limited to a few lattice points, superlattices etc. but with the constraint that the system is initially not in a pure state. Even for extended systems such as the one treated in the recently reported polaron study \cite{Gaal2007}, a similar non linear mixing of frequencies might be observable.

\end{document}